\documentclass{ws-procs975x65}

\def\be {\begin{equation}}
\def\ee {\end{equation}}
\def\ba {\begin{eqnarray}}
\def\ea {\end{eqnarray}}
\def\nn {\nonumber}
\def\a  {\alpha}
\def\b  {\beta}

\def\d  {\delta}
\def\D  {\Delta}
\def\e  {\epsilon}

\def\o  {\omega}

\def\la {\label}
\def\le {\left}
\def\ri {\right}

\def\bi {\begin{itemize}}
\def\ei {\end{itemize}}

\def\vs {\vspace}

\def\bc {\begin{center}}
\def\ec {\end{center}}

\begin{document}

\title{
The Generalized Uncertainty Principle and Quantum Gravity Phenomenology
}

\author{Ahmed Farag Ali, Saurya Das}

\address{
Dept. of Physics,
University of Lethbridge, 4401 University Drive, \\
Lethbridge, Alberta, Canada T1K 3M4 \\
%
%University Department, University Name,\\
%City, State ZIP/Zone, Country\\
E-mails: ahmed.ali@uleth.ca, saurya.das@uleth.ca \\
http://directory.uleth.ca/users/ahmed.ali, 
http://people.uleth.ca/$\sim$saurya.das }

\author{Elias C. Vagenas}

\address{
Research Center for Astronomy \& Applied Mathematics,\\
Academy of Athens, \\
Soranou Efessiou 4, GR-11527, Athens, Greece \\
%Group, Laboratory, Street,\\
%City, State ZIP/Zone, Country\\
E-mail: evagenas@academyofathens.gr \\
http://users.uoa.gr/$\sim$evagenas
}

\begin{abstract}
In this article we examine a Generalized Uncertainty Principle
which differs from the Heisenberg Uncertainty Principle by terms linear and quadratic
in particle momenta, as proposed by the authors in an earlier paper. We show
that this affects all Hamiltonians, and in particular those
which describe low energy experiments. We discuss possible observational consequences.
Further, we also show that this indicates that space may be discrete at the fundamental
level.
\end{abstract}


%\bodymatter

%\section{Introduction}
%\label{intro}

\vs{.4cm}

String Theory \cite{acv}, certain other approaches to Quantum Gravity, as well as
Black Hole Physics \cite{maggiore} suggest a modification of the
Heisenberg's Uncertainty Principle near the Planck scale to a so-called Generalized Uncertainty
Principle (GUP) of the form
\be
\Delta p~\Delta x \geq \frac{\hbar}{2}
\left[
1 + \beta_0 \frac{\ell_{Pl}^2}{\hbar^2} \Delta p^2
\right]
\la{gup1}
\ee
where
$\ell_{Pl} = \sqrt{\frac{G\hbar}{c^3}} = 10^{-35}m$ is the Planck length and
$\b_0$ is a constant, normally assumed to be of order unity. Evidently, the
new second term on the RHS of (\ref{gup1}) is important only when $x,\D x\approx \ell_{Pl}$ or
$p, \D p \approx p_{Pl} \approx 10^{16} TeV/c $ (the Planck momentum),
i.e. at very high energies/small length scales.
Inverting Eq.(\ref{gup1}), we get
$ \Delta p \leq \frac{\hbar}{\beta_0\ell_{Pl}^2} \left[ \Delta x
\pm \sqrt{\Delta x^2 - \beta_0\ell_{Pl}^2} \right]$,
implying the existence of a {\it minimum} measurable length
$\Delta x \geq \Delta x_{min} \equiv \sqrt{\beta_0} \ell_{Pl}$.
It can be shown that the above GUP can be derived from a modified
Heisenberg algebra \cite{kmm}
\be
[x_i, p_j] = i \hbar [ \d_{ij} + { \frac{\beta_0 \ell_{Pl}^2}{\hbar^2} (p^2 \delta_{ij} + 2 p_i p_j)}]~.
\la{algebra1}
\ee
On the other hand, Doubly Special Relativity (DSR) theories \cite{dsr}
suggest yet another modified
algebra between position and momenta \cite{cg}
\be
[x_i,p_j] =
i \hbar [ (1-\ell_{Pl} |\vec p|)\delta_{ij} + \ell_{Pl}^2 p_i p_j ]
\la{algebra2}
\ee
as well as the existence of a {\it maximum} observable momentum
$\Delta p \leq \Delta p_{max} \approx M_{Pl}c$.
Using the Jacobi identity
$
\left[ [x_i,x_j],p_k \right] +
 \left[ [x_j,p_k],x_i \right] +
 \left[ [p_k,x_i],x_j \right]  = 0
$
and the assumption that space commutes with space and momenta with momenta,
algebras (\ref{algebra1}) and (\ref{algebra2}) can be reconciled as limits of
a single algebra of the form \cite{dv3}
\footnote{We also cite reference [6] for more references to earlier works.}
\be
[x_i,p_j] = i\hbar \left[
\delta_{ij} - {\alpha \left( p \delta_{ij} + \frac{p_i p_j}{p} \right)
+ \alpha^2 (p^2 \delta_{ij} + 3 p_i p_j )}
\right]~.
\ee
Here $\alpha = \frac{\alpha_0}{M_{Pl} c} = \frac{\alpha_0 \ell_{Pl}}{\hbar}$.
Again, $\alpha_0$ is normally assumed to be of order unity.
The above algebra predict both a $\Delta x_{min}$ and a $\Delta p_{max}$. It also implies the
following representation of the momentum operator in position space
$p_j = p_{0j} \left( 1 {- \alpha p_0 + 2\alpha^2 p_0^2 }\right)$
where
$p_{0j} = -i\hbar\frac{\partial}{\partial x_j}$ is canonical (but unphysical) and satisfies the usual
commutator $[x_i,p_{0j}]=i\hbar \delta_{ij}$.
Correspondingly, a non-relativistic Hamiltonian takes the form
$ H =
\frac{p^2}{2m} + V(\vec r)
=  \frac{p_0^2}{2m} + V(\vec r) - \frac{i\hbar^3\alpha}{m}\frac{d^3}{d x^3}
$
where the last term can be considered as a Quantum Gravity induced perturbation in
the time-dependent Schr\"odinger Equation
\ba
\le[ H_0 {+ H_1} \ri]\hspace{-0.3ex}\psi
= \le[-\frac{\hbar^2}{2m} \frac{d^2}{dx^2} +V(x)
- {i \frac{\alpha\hbar^3}{m} \frac{d^3}{dx^3}}
\ri]\hspace{-0.3ex}\psi
= i\hbar \frac{\partial \psi}{\partial t}~.\nn
\la{se1}
\ea
The above equation admits of a new conserved current
$
J = \frac{\hbar}{2mi} \le( \psi^\star \frac{d\psi}{dx}
- \psi \frac{d\psi^\star}{dx}
\ri) + { \frac{\alpha \hbar^2}{m}
\le(
\frac{d^2|\psi|^2}{dx^2} - 3 \frac{d\psi}{dx} \frac{d\psi^\star}{dx}
\ri)}
$
and charge $\rho = |\psi|^2~$, such that
$\frac{\partial J}{\partial x} + \frac{\partial \rho}{\partial t} = 0 $.
The effect of the perturbation can be found for example on a simple harmonic oscillator,
with $V=m\omega^2x^2/2$, for which the shift in the ground state energy eigenvalues is,
using second order perturbation theory
$
\frac{\Delta E_{GUP(0)}}{E_0} \sim {\hbar \omega m \a^2}~.
$

Concerning Landau Levels, for a particle of mass $m$,
charge $e$ in a constant magnetic field ${\vec B} = B {\hat z}\approx~10T$, ${\vec A}=Bx {\hat y}$
and cyclotron frequency $\o_c=eB/m$,
the Hamiltonian is
$
H = \frac{1}{2m}\le( \vec p_0 - e \vec A\ri)^2
{- \frac{\a}{m}\le( \vec p_0 - e \vec A\ri)^3}
= H_0 { - \sqrt{8 m } \a H_0^{\frac{3}{2}} }
$
and the energy shifts are
$
\frac{\Delta E_{n(GUP)}}{E_n} = { -\sqrt{8 m} \a  (\hbar \omega_c )^{\frac{1}{2}}
 (n +\frac{1}{2})^{\frac{1}{2}} } \approx { - 10^{-27} \a_0 }~,
$
from which we conclude that if
$\a_0 \sim 1$, then $\frac{\Delta E_{n(GUP)}}{E_n}$ is too small to measure.
On the other hand, with
current measurement accuracy of $1$ in $10^3$, one obtains the
following upper bound on the GUP parameter: $\a_0 < 10^{24}$.

Similarly for a Hydrogen atom with standard Hamiltonian
$H_0 = \frac{p_0^2}{2m} - \frac{k}{r}~$ and perturbing Hamiltonian $H_1 = -\frac{\a}{m} p_0^3$,
it can be shown that the GUP effect on the Lamb Shift is
$
\frac{\D E_{n(GUP)}}{\D E_n} = 2 \frac{\D |\psi_{nlm(0)}| }{\psi_{nlm}(0)}
\approx  \a_0 \frac{4.2 \times 10^4 E_0}{27 M_{Pl C^2}}
\approx 10^{-24}~\a_0
$~.
Again, if $\a_0 \sim 1$,  then $\frac{\Delta E_{n(GUP)}}{E_n}$ is too small, whereas
with current measurement accuracy of $1$ in $10^{12}$, we infer
$\a_0 < 10^{12}$. For some other examples, we refer the reader to our earlier papers \cite{dv1,dv2}~.

%%%%%%%%%%%%%%%%%%%%
%
% PARTICLE IN A BOX
%%%%%%%%%%%%%%%%%%%%%

Finally, we consider the free-particle Schr\"odinger equation for a particle in a box of length $L$
\cite{dv3},
with the solution $\psi(x) = {A} e^{ik'x} + B e^{-ik''x} + { C e^{\frac{ix}{2\a\hbar}}}$.
Note the appearance of a {\it new} oscillatory term.
Here $k' = k(1 {+k\a\hbar})~,~k'' = k(1{-k\a\hbar})$ (to leading order in $\a$).
The boundary condition $\psi(0)=0$ implies $A + B + C = 0$, and in addition to
the boundary condition $\psi(L)=0$, this yields
\ba
2iA \sin(kL) &=& {\left|C\right|
\le[
e^{-i(kL+\theta_C)}
-e^{i(L/2\a\hbar - \theta_C)}
\ri]}
+ {\cal O}(\a^2)~,
\ea
where $ C = |C|e^{-i\theta_C} $. Taking real parts of both sides
(assuming $A$ is real, without loss of generality), we get
$
{\cos\!\le(\frac{L}{2\a\hbar} -\theta_C \ri) = \cos(kL+\theta_C) = \cos(n\pi + \theta_C +\e)}
$
which has the solutions
\ba
\frac{L}{2\alpha\hbar}  = \frac{L}{2\a_0 \ell_{Pl}} = n\pi + 2q \pi + 2\theta_C
~\mbox{or}~
= - n\pi + 2q\pi
%{\frac{L}{2\alpha\hbar} }
%=  \frac{L}{2\a_0 \ell_{Pl}}
~~~[n,q \in \mathbb{N}]~.
\ea
From the above we conclude that a particle can be confined only in boxes of certain
discrete lengths, and further speculate that this might indicate that all measurable lengths are
quantized, since measurement of lengths require at least one particle, possibly many.
We think that this result can be generalized to relativistic particles, as well as to
the quantization of areas and volumes \cite{dv4}~.

In summary, in this article we have shown that a single GUP exists, which is consistent with the
predictions of Black Hole Physics, String Theory, DSR etc., and that this induces
perturbations to all Hamiltonians. Applying this to a few concrete examples such as the
Harmonic Oscillator, Landau Levels and Lamb Shift, we have computed corrections
due to this perturbation. From these, we concluded that if the GUP parameter
$\a_0$ is of order unity, these corrections are probably too small to be measured
at present. On the other hand, current experimental accuracies impose upper bounds
on the GUP parameter. Finally, by solving the GUP corrected Schr\"odinger equation
for a particle in a box, we have shown that boundary conditions require
the box length to be quantized, suggesting quantization of measurable lengths, and
possibly of surfaces and volumes as well. We hope to report further on these elsewhere.

This work is supported by the Natural Sciences and Engineering Research Council
of Canada and the Perimeter Institute for Theoretical Physics.

\vs{-.1cm}

%\bibliographystyle{ws-procs975x65}
%\bibliography{ws-pro-sample}

\end{document}